# The Influence of Neutron Irradiation on $(B_{0.65}C_{0.35})Ba_{1.4}Sr_{0.6}Ca_2Cu_3O_z$ Superconducting Phase: the Role of the Grain Edge.

V. Mihalache[1], A. Totovana[1], V. Sandu[1], S. Popa[1], G. Aldica[1], A. Iyo[2,3]

[1] National Institute of Materials Physics, Bucharest-Magurele, POB-MG-7, R-76900 Romania

[2] Nanoelectronics Research Institute of National Institute for Advanced Industrial Science and Technology, 1-1-1 Umezono, Tsukuba, Ibaraki, 305-8568 Japan

[3] CREST JST, Kawaguchi, Saitama, 332-0012 Japan

# The Influence of Neutron Irradiation on $(B_{0.65}C_{0.35})Ba_{1.4}Sr_{0.6}Ca_2Cu_3O_z$ Superconducting Phase: the Role of the Grain Edge.

V. Mihalache[1], A. Totovana[1], V. Sandu[1], S. Popa[1], G. Aldica[1], A. Iyo[2,3]

[1] National Institute of Materials Physics, Bucharest-Magurele, POB-MG-7, R-76900 Romania

[2] Nanoelectronics Research Institute of National Institute for Advanced Industrial Science and Technology, 1-1-1 Umezono, Tsukuba, Ibaraki, 305-8568 Japan

[3] CREST JST, Kawaguchi, Saitama, 332-0012 Japan

**Abstract**

Using the transport and magnetization measurements the influence of neutron irradiation at a fluence of $5 \cdot 10^{17}$ n·cm$^{-2}$ on $(B_{0.65}C_{0.35})Ba_{1.4}Sr_{0.6}Ca_2Cu_3O_z$ has been investigated. The neutron irradiation was found to decrease critical temperature and transport critical current density, increase the residual and normal state resistivity, and improve the intragranular critical current density with $1.6 \times 10^5$ A/cm$^2$ (at 77.3K and in the applied field up to 160 kA.m) and $\Delta M_{irr}/\Delta M_{nonirr}$ ratio (up to factor of 3) at highest field used for investigation. The field dependence of this ratio, which is below the unity at very low field but higher than 1 at high fields, correlated with the shape of the histertic loops as well as with the change of the transport parameters after irradiation suggests the role of the irradiation induced effects on the grain edges. We discuss these effects in the framework of the Bean-Livingstone surface barriers and geometrical barriers.

**Introduction**

Vortex dynamics in HTS is very sensitive to thermal fluctuations due to the very small superconducting coherence length and the large anisotropy resulting from their laminar structure. The large influence of the thermally activated depinning is the main reason of the current limitation reflected in the fast drop of the critical current density, $J_c$, with increasing temperature. Therefore, the increase of the pinning potential in HTS is crucial for applications.

One of the most efficient ways to improve the flux pinning capability consists in exposing these materials to a suitable radiation environment. The enhancement of flux pinning after neutron irradiation was obtained in different types of HTS materials as single crystals ($YBa_2Cu_3O_{7-\delta}$, $Bi_2Sr_2CaCu_2O_8$, and $Tl_2Ba_2Ca_2Cu_3O_y$) [1-3], pollycrystals ($YBa_2Cu_3O_{7-\delta}$, [4-6], ($Cu_{1-x}C_x$)$Ba_2Ca_3Cu_4O_{10+\delta}$ and (Cu,C)- ($Cu_{1-x}C_x$)$Ba_2Ca_4Cu_5O_{10+\delta}$ [7] $Bi_2Sr_2Ca_2Cu_3O_{108}$ [8.9], and as Ag/Bi-2223 tapes [10]. In pollycrystals, the increase of the intragranular critical current density cannot be dissociated from some side effects on the processes controlled by the grain geometry and interfaces. Less discussed, these effects prove to be, however, crucial for the current transport in polycrystalline HTS.

Among the HTC superconductors, $(B_{0.65}C_{0.35})Ba_{1.4}Sr_{0.6}Ca_2Cu_3O_z$ is of special interest for two reasons; first the high boron content, which is the element with highest effective cross section for the neutron capture ($\sigma = 3837$ barn for the isotope [10]B which constitute 19.9% in natural boron), makes possible a decrease of the irradiation fluences comparative with other HTS materials, and, second, the fission reaction $B(n,\alpha)Li$ products deposit much more energy in the sample as compared with the energy resulting from fission of [6]Li, which usually is inserted in HTS in order to increase the irradiation efficiency. By our knowledge there is no report concerning the effects of neutron irradiation in these class of HTS.

In the this work we focus on the effect of neutron irradiation on the superconducting properties of sintered $(B_{0.65}C_{0.35})Ba_{1.4}Sr_{0.6}Ca_2Cu_3O_z$ (BC-1223) samples with an emphasis on the intergrain region which is supposed to be the "valve" controlling the transport critical current density. We analyze the sources of the change in the irreversibility connected to the surface (bean-Livingstone) and geometrical barriers.

**Experimental.**

Nearly single-phase BC-1223 samples were synthesized from starting powders of precursors (Ba Ca Cu O and Sr Ca Cu O), $B_2 O_3$ and Ag $CO_3$. The latter was used only as a $CO_3$ source. The precursors, $Ba_2 Ca_2 Cu_3 O_y$ and $Sr_2 Ca_2 Cu_3 O_z$, respectively, were prepared by calcining a well ground mixture of $BaCO_3$ (or $SrCO$), $CaCO_3$ and CuO powders in a flowing oxygen atmosphere. The starting mixture was sealed in a gold capsule and heated at 1100° C for 2 h under a pressure of 4.5 GPa using a cubic-anvil-type high-pressure apparatus. More detalies on the sample preparation are given in [11,12]. The powder X-ray diffraction patterns (Fig. 1) show a nearly single BC-1223 phase. According to the sample morphology studies by scanning electronic microscopy (SEM) the mean grain size of the platelets is 5 μm. The electrical resistance was measured by the standard four-probe method and the critical temperature was taken at the inflexion point of the superconducting transition. Zero-field transport critical current density was extracted from the

current–voltage curve using the criterion of 1μV/cm. The small field DC magnetization measurements were performed at liquid nitrogen temperature, up to a field value of 150 kA/m, using an integrator magnetometer (Thor Cryogenics 9020II). The field was applied on the largest face of the sample, i.e. in the direction of the pressing force. The sample was cut up into two pieces, one designated to be irradiated and one kept as reference. The irradiation was performed in the hot chamber of the nuclear reactor TRIGA 2 at a fluence of $5 \times 10^{17} cm^{-2}$. The samples were introduced in aluminum blocks which were suspended in the center of one channel inside. During the irradiation, the temperature in the aluminum blocks was not measured, but the channel temperature was not higher than 40 °C. After irradiation, the samples were maintained in the hot chamber for 7 days in order to remove the residual activity.

**Results**

Figure 2 shows the temperature dependence of the sample resistance before and after irradiation and  the inset shows the temperature dependence of the reduced resistance. As expected the critical temperature shows an important depression from 116.6 K to 100 K and the normal state resistance increases as well as the transition width (approximately two times). The linear in $T$ dependence of the resistivity, which is conspicuous in the virgin sample above the Gaussian fluctuation controlled range, gets a complicated dependence in the irradiated sample. Because it less likely a loss of oxygen during the irradiation process, the main reason of this behavior resides in the defects introduced by irradiation. Indeed, neutron diffraction studies on $YBa_2Cu_3O_{6.95}$ [13] have shown a certain disorder in oxygen sublattices, mainly in the O(4) and O(5) sites. Additionally, the Debye–Waller factor, which reflects dynamic and static atomic shifts from the regular sites, has been reported to increase with fluence slowly. The fact that small disorder can be effective to depress superconductivity with almost 16 K was explained [14] by the closeness of HTS materials to Anderson localization due to their quasi two-dimensionality. However, superconductivity should survive to localization as long as the latter is weak enough so that the localization length be longer than the superconducting coherence length.

There is also another effect, which can explain the robustness of superconductivity, which results from the kinetics of defects. Usually, they accumulate by different kind of pre-irradiative defects (sinks) and, in polycrystalline samples, grain border is a very attractive region toward which defects migrate and accumulate. Hence, the disorder is mainly transferred to the border region increasing the intergrain resistivity, which now controls the flowing of current. This fact is reflected in the severe decrease of the transport critical current density ($J_{ct}$) from 130 A/cm$^2$ to 5 A/cm$^2$. Similar effects were reported by Sen *et al.* [15] on proton irradiated Bi-based polycrystalline superconductors. They found up that the irradiation-induced defects accommodate in the grain boundary region causing an increase in junction barrier thickness.

The most striking change following irradiation was observed in the applied field $H_a$ dependence of the magnetization $M$. As a general feature, both irradiated and unirradiated samples exhibit a relatively weak pinning but with a conspicuous, almost three times, enhancement for the irradiated sample. However, shows an important difference between the shape of $M$ vs $H_a$ loop of the irradiated sample as compared with the nonirradiated sample (Fig. 3) and also in the dependence of the irreversible magnetization $\Delta M = M^+ - M^-$ (Fig.4) with M$^-$ and M$^+$ the descending and the ascending magnetization, respectively. The absolute value of the entry magnetization peak at very low applied field is seriously suppressed in the irradiated sample and shifted to lower applied fields, but the decrease of the ascending branch $M$ with increasing field is very slow and, at a certain field, it overtops the magnetization of the unirradiated sample. Even having the same sign, the descending branch $M^+$ is much closer to zero magnetization. As a result, the intragranular critical current density $J_{cg}$, which is connected to the irreversible magnetization by

Bean relation, $J_{cg} = \dfrac{3\left(M^- - M^+\right)}{2R_g}$ , becomes dominant for $H_a > 70$ kA/m and increases up to $J_{cg} \approx 1.6\times10^5$ A/cm$^2$, almost three times, at $H_a \approx 150$ kA/m as compared with the unirradiated sample (see inset to Fig 4). Here $2R_g$ is the average grain size. Hereafter, we will label with the subscripts (0) and (irr) the unirradiated and irradiated characteristics, respectively. The intragranular critical current density diminishes exponentially with increasing field in the unirradiated sample as

$J_{cg,0}(H_a) = 1.1\times10^5 \exp\left(-\dfrac{H_a}{H_{0,0}}\right)$ (kA/m$^2$) with $H_{0,0} \cong 66$ kA/m. The irradiated sample has a much slower, linear decrease; $J_{c.irr}(H_a) \approx 4.22\times10^5\left(1 - \dfrac{H_a}{H_{0,irr}}\right)$ with H$_{0,irr}$ = 426 kA/m. For the analysis of transport process it is important to inspect the hysteresis loops for very low applied field, $H_a < 15$ kA/m. This should allow an estimate of the main characteristic fields governing the intergrain region as well as their variation under irradiation. At extremely low field, both samples, unirradiated and irradiated, exhibit a linear dependence of the virgin magnetization vs applied field $H_a$ up to certain field $H_{c1j}$ where it deviates from linearity. A further increase of the applied field leads to a new linear $M$ vs $H_a$ dependence above a field $H_{c2j}$. The latter dependence ends up at a field $H_{c1g}$ where a second deviation from linearity is conspicuous. (see Fig. 5a and 5b).

**Discussion**

Both the transport current and the low field behavior of the magnetization can be understood starting from the sample morphology which is a collection of superconducting grains separated by nonsuperconducting boundaries acting as weak-links. This weak links allows the flow of a Josephson supercurrent as long as the field is low enough to prevent its vanishing. Therefore, the characteristic fields $H_{c1j}$ and $H_{c2j}$ observed in very low fields on the virgin magnetization curve are identified with the critical Josephson fields, whereas $H_{c1g}$ marks the penetration of vortices within superconducting grains. The transport critical current can be identified with the average Josephson current and is strongly dependent on the average junction thickness $d$. All these fields are strongly reduced by the irradiation process. $H_{c1j}$ decreases from 1.29 to 0.4 kA/m; $H_{c2j}$ diminishes from 3.57 to 0.79 kA/m, and $H_{c1g}$ is reduced from 11.8 to 3.0 kA/m. The remnant magnetization is also reduced from 0.74 to 0.1 kA/m.

The value of the slope of the magnetization below $H_{c1j}$ is almost halved as compared with the perfect shielding proposed by different theoretical models [16-19]. This deviation arises from the nature of the sintering process which leaves behind cavities, voids, thick intergrain connections unable to carry supercurrent etc. For this reason we avoided this part of magnetization for quantitative evaluations. When $H_a > H_{c1j}$, Josephson vortex lines penetrate into the sample along the grain boundaries where they become more or less pinned and determine the field gradient able to sustain an intergrain current. Increasing field they accumulate and at $H_{c2j}$ the current density cancels as well as its contribution to the magnetization. Therefore, the intergranular critical current density could be obtained from $H_{c2j}$ [16], the field where the magnetization of the intergrain currents vanishes, as

$$J_{cr,\mathrm{int}\,er} \propto \frac{H_{c2j}}{r_p} f(N) \tag{1}$$

Here $r_p$ is the range of the pinning potential and $N$ the demagnetizing factor. A rough approximation with $r_{p,0} \cong r_{p,\mathrm{irr}}$ and $N_0 \cong N_{\mathrm{irr}}$ would give $\dfrac{J_{cr,\mathrm{int}\,er}^0}{J_{cr,\mathrm{int}\,er}^{irr}} \approx 5$ which is more than five times smaller than the critical field obtained from transport measurements, $\dfrac{J_{ct}^0}{J_{ct}^{irr}} \approx 26$. Therefore, we have to consider all the term in Eq. 1 as contributing to the change in transport properties.

Useful information can be provided by the slope of the second linear range of the virgin magnetization which is connected both to the effective permeability $\mu_{eff}$ and to the demagnetizing factor $N$ [3]:

$$-\frac{dM_{vir}}{dH} \propto \frac{1-\mu_{eff}}{1-\left(1-\mu_{eff}\right)N} \quad \text{for } H_{c2j} < H < H_{c1g} \tag{2}$$

In the latter field range, the linear dependences of the virgin magnetization have the following particular parameters:

$$-M_{vir,0} = 0.254H + 0.095$$

$$-M_{vir,irr} = 0.317H + 0.078$$

for the unirradiated and irradiated sample, respectively.

The increase of the slope in the irradiated sample should suggests, if $N \ll 1$, a surprising improvement of the superconducting properties of the grains. Indeed, from $1-\mu_{eff} \approx f_s\left(1-\dfrac{\lambda_g}{2R_g}\right)$ and $\lambda_g \propto f_s^{-1}$, with $f_s$, the weight of the superconducting contribution in the whole sample, it results an increase of the fraction of superconducting contribution. This is confirmed also by the intragranular irreversibility, which suggests the strengthening of the bulk pinning; hence the superconducting condensation energy is at least as high as in the case of unirradiated sample. The enhancement of the bulk properties is the result of the kinetics of the defect generation, which supposes both generation and irradiation-stimulated recombination of defects. Which of these two processes is dominant depends on the concentration of the pre-irradiation defects. Once created, the defects are grabbed by some of the pre-irradiative ones, among which the most important is the grain edge. In fact, the grain border is constituted from a high density of defects, mainly disorder in the oxygen sublattice, which suppress the superconducting order parameter and give rise to the weak-link [19, 20]. Therefore, it is a high probability that the new created defects should migrate toward the grain edge, unless they are not collected by bulk defects, and increase the effective thickness $d$ of the intergrain space. This generates a decrease of both critical Josephson current density, $j_c \propto \exp\left(-\dfrac{d}{\xi}\right)$ and the lower Josephson critical field, $H_{c1j} \propto d^{-1/2}$. The accumulation of the defects at the weak link might be also responsible for the differences observed in transport measurements ($T_c$ and $j_{ct}$). For example, a comparison of the critical current densities obtained from transport and magnetic measurements can provide an (over)estimate of $r_{p, irr}$ to compensate the smaller increase in $H_{c2j}$, namely $r_{p,irr} = 6\ r_{p.0}$. On the other side, the clusterization of both new and some of the pre-irradiated defects in strong pinning centers and at the grain surface, leaves behind much cleaner, almost defect-less, areas which would explain the increase of $f_s$. A question still remains: which is the reason of the decrease of the lower intragranular critical field $H_{g1}$ weather $\lambda_g$ decreases for it is well established that $H_{c1g} \propto \lambda_g^{-2}$ ?

The answer can be found considering the presence of two kinds of effects that increase the threshold for field penetration. One arises from the surface barriers of the Bean-Livingston (BL) type [21] and for the polycrystalline sample is as important as it is in the case of single crystals [22, 23]. The source of this barrier is the attraction of the Abrikosov vortex to its "mirror image" near the surface. The presence of this surface effect prevents flux penetration at lower critical field, at which internal flux tread become energetically favorable, until a higher thermodynamic critical field ($H_c$) is reached. A fingerprint of the surface barriers is thought to be the asymmetric shape of the hysteretic loop, namely the sharp drop-off of the magnetization above the flux penetration field $H_p$ on the ascending branch and the nearly flatness of the descending branch. Fig. 3 shows that the shape of magnetization loop of nonirradiated sample could mirror the presence of BL barrier. It was believed that the roughness of the barrier is crucial for the penetration of vortices. However, Bass *et al.* [24] have shown that even high roughness reduces only with 10% the surface barrier. Much more important are the weak points at the grain edge with depressed order parameter where the critical energy for vortex nucleation is diminished. Since the irradiation introduces defects into the surface, the BL surface barrier is depressed [25] and become less important for the pinning. This is consistent with our results, which shows an almost four times reduction of $H_{c1g}$ after irradiation. The mechanism is similar with that one proposed by Koshelev and Vinokur [26] in which the strong pinning centers located in the proximity of the grain surface acts as weak spots and facilitate the creep of the vortices over the surface. The creep occurs either by direct interaction vortex–pin or via the disturbed Meissner current. Subsequently, the vortices can be transferred to other deeper defects. An attempt to fit the entry magnetization below $H_p$ suggested a power low type dependence for the unirradiated sample, $M_{en,0} \propto H^{-0.75}$ whereas the irradiated sample exhibits a much slower decrease $M_{en,irr} \propto H^{-0.16}$. It is worthy to notice that the exponent of the unirradiated sample is close to the value predicted for surface barriers $\alpha = -1$ [27].

The second effect that can intervene in the field behavior arises from the geometrical barriers [28]. The geometrical barrier is a result of nonellipsoidal sample geometry and is stronger for specimanes with a high aspect ratio that produce an increase of the line tension at the graine edge.

Once the first vortices have penetreted, the Lorentz force of the Meissner current drives them to the centre of the sample where they concentrate in increasing field (in the absence of the pinning centers). With decreasing field, the flux remain trapped by the same shielding currents expanding of the profile of the vortices to fill the specimen. This process leads to strong hystersis in the magnetization even in the complete absence of pinning. Like surface BL barrier, a geometrical barriers results in a delayed initial penetration field. Besides, the magnetic hysterezis loop is strongly asymmetric. However, geometrical barrier acts on a macroscopic scale whereas surface barriers acts on a microscopic scale and is a subject to thermal activation. Therefore, the geometrical barrier are strongly dependent on the geometry of the sample which are not expected to support dramatic changes during irradiation. Deffect accumulation is important only for the weak links which are very narow as compared with the grain size. Nevertheless, some grains from strongly coupled clusters that behaves like large single grains. Irradiation might disconnect them by the addition of new defects intracluster boundaries, hence to reduce the geometrical barrieres by reducing the aspect ratio. We did not estimate this effect in our case, but the imperfect shielding for $H_a < H_{c1j}$ (the slope of $M / H_a < 1$) suggests that it is negligible.

**Conclusions.**

In conclusion the influence of neutron irradiation at a fluence of $5 \times 10^{17}$ n cm$^{-2}$ on $(B_{0.65}C_{0.35})Ba_{1.4}Sr_{0.6}Ca_2Cu_3O_z$ was studied. The temperature at which the resistance become zero $T_c^{R=0}$ (critical temperature) has decreased; residual resistance $R_0$ and normal state resistance have increased after irradiation. Irradiation leaded to significant decrease of transport critical current density ($J_{ct}$) from 130 A/cm$^2$ to 5 A/cm$^2$. The significant increasing of $\Delta M_{irr}/\Delta M_{nonirr}$ (H) ratio (up to a factor of 3) is observed for irradiated sample at fields above 70kA/m. Increasing of critical current density in the grains, $\Delta J_{cg} = J_{cg}^{irrad} - J_{cg}^{nonirr}$, after irradiation is up to $1.6 \times 10^5$ A/cm$^2$. The decreasing $\Delta M_{irr}/\Delta M_{nonirr}$ at lower fields, the change in the shape of the hysteretic loops, as well as the change of the transport parameters were explained based on the irradiation induced modifications of the grain edges, which, in turn, dramatically modify the BL surface barriers.

**References.**

**Figure Caption**

**Fig. 1**. X-Ray difractogram of the nonirradiated sample. A nearly single BC-1223 phase is present.

**Fig. 2.** Temperature dependence of the sample resistance before and after irradiation. The critical temperature was depressed from 116.6 to 100 K.

**Fig. 3**. Magnetization $M$ vs applied field $H_a$ plots for both virgin and irradiated samples. The absolute value of the entry magnetization peak is suppressed in the irradiated sample and shifted to lower $H_a$ whereas the ascending branch $M$ decreases slowly and, at a certain field, overtops the magnetization of the unirradiated sample.

**Fig. 4**. The dependence of the irreversible magnetization $\Delta M$ on the applied field $H_a$, before and after irradiation. $\Delta M_{irr}$ becomes higher than $\Delta M_{virg}$ above $H_a > 70$ kA/m. Inset: field dependece of the intragranular critical current density $J_{cg}$. $J_{cg}$ (bulk pinning) becomes dominant increases up to $J_{cg} \approx 1.6 \times 10^5$ A/cm$^2$ at $H_a \approx 150$ kA/m as compared with the nonirradiated sample.

**Fig. 5**. The magnetization at very low applied fields $H_a \leq 13$ kA/m; a) nonirradiated sample with the characteristic fields: $H_{c1j} = 1.29$ kA/m, $H_{c2j.} = 3.57$ kA/m, and $H_{c1g} = 11.8$ kA/m; b) irradiated sample with the characteristic fields: $H_{c1j} = 0.4$ kA/m, $H_{c2j.} = 0.79$ kA/m, $H_{c1g} = 3.0$ kA/mkA/m.

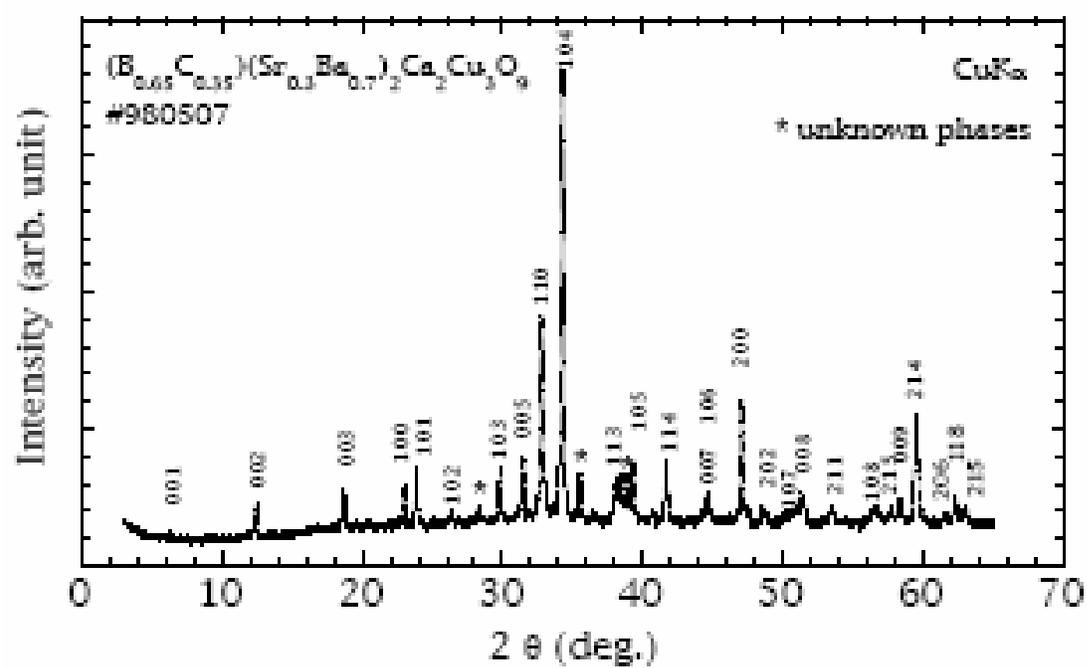

Fig. 1

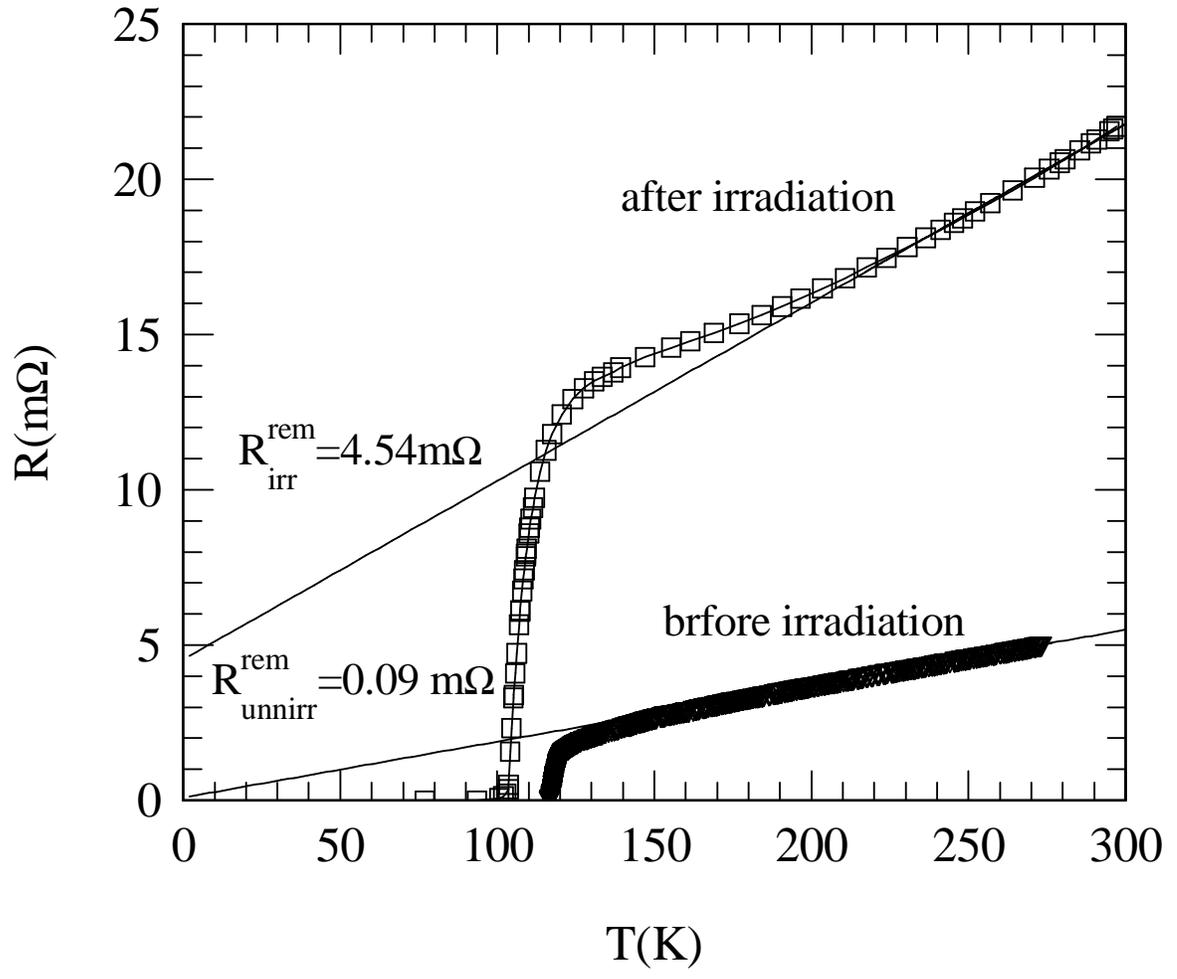

Fig. 2

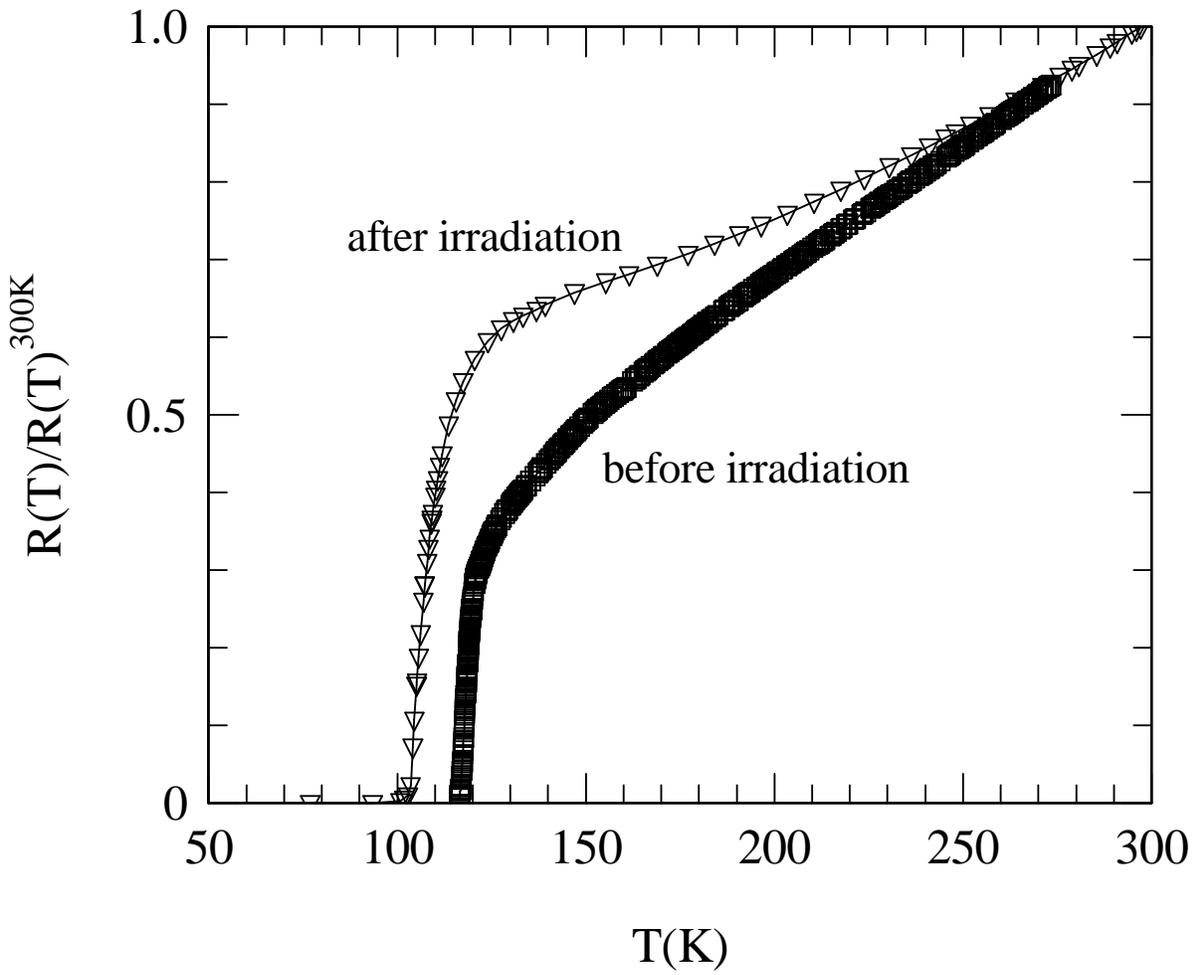

Inset to Fig 2

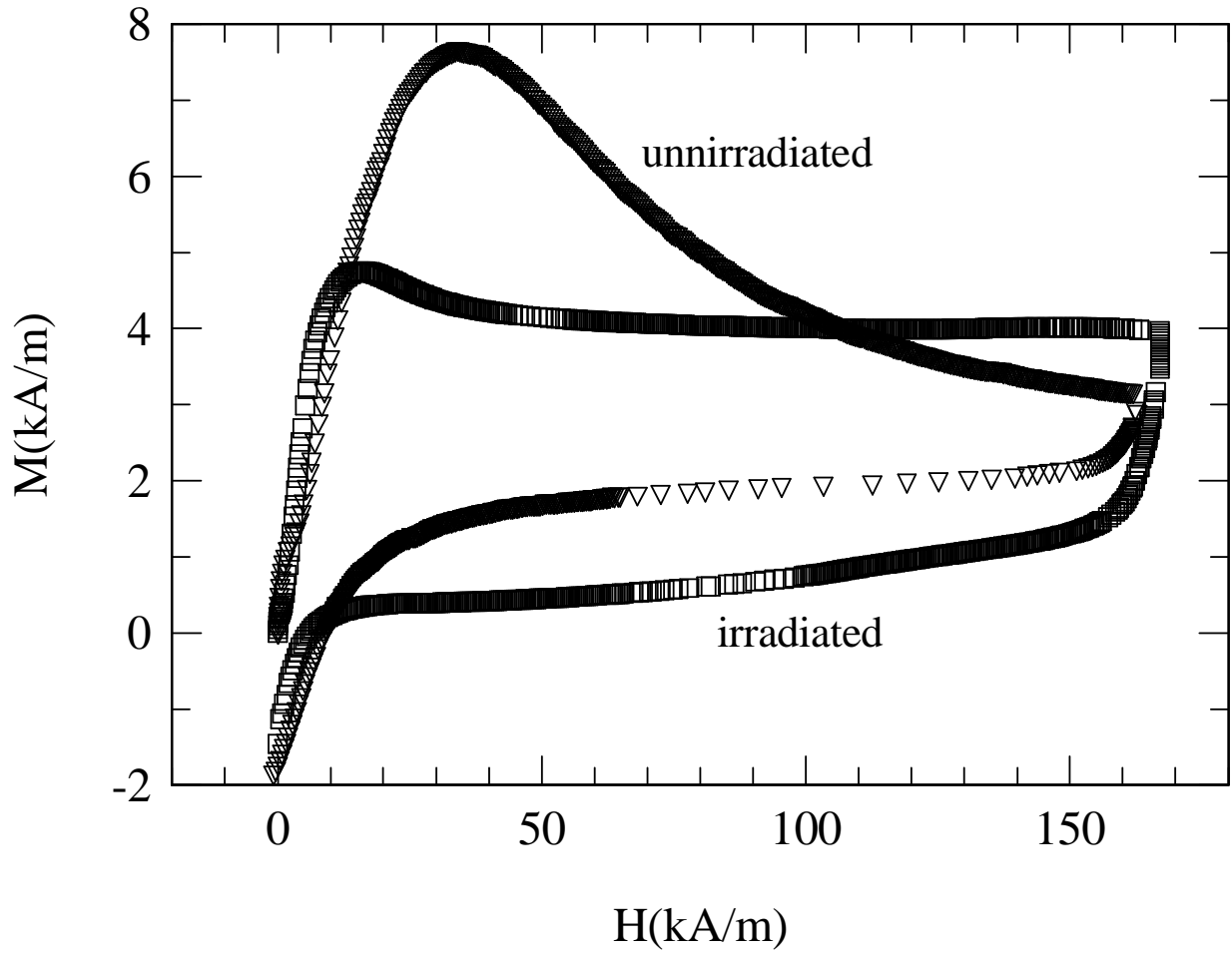

Fig. 3

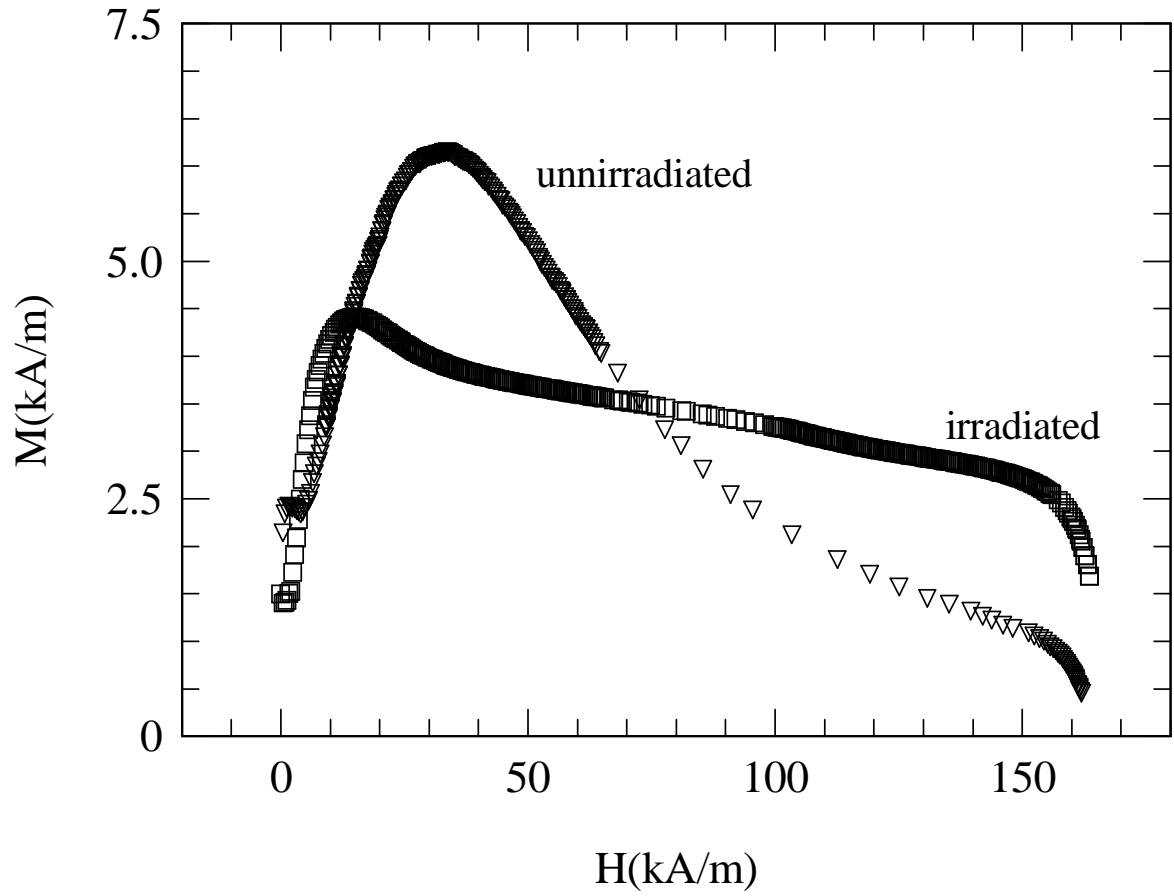

Proba BC-$^{18}$1223(1)

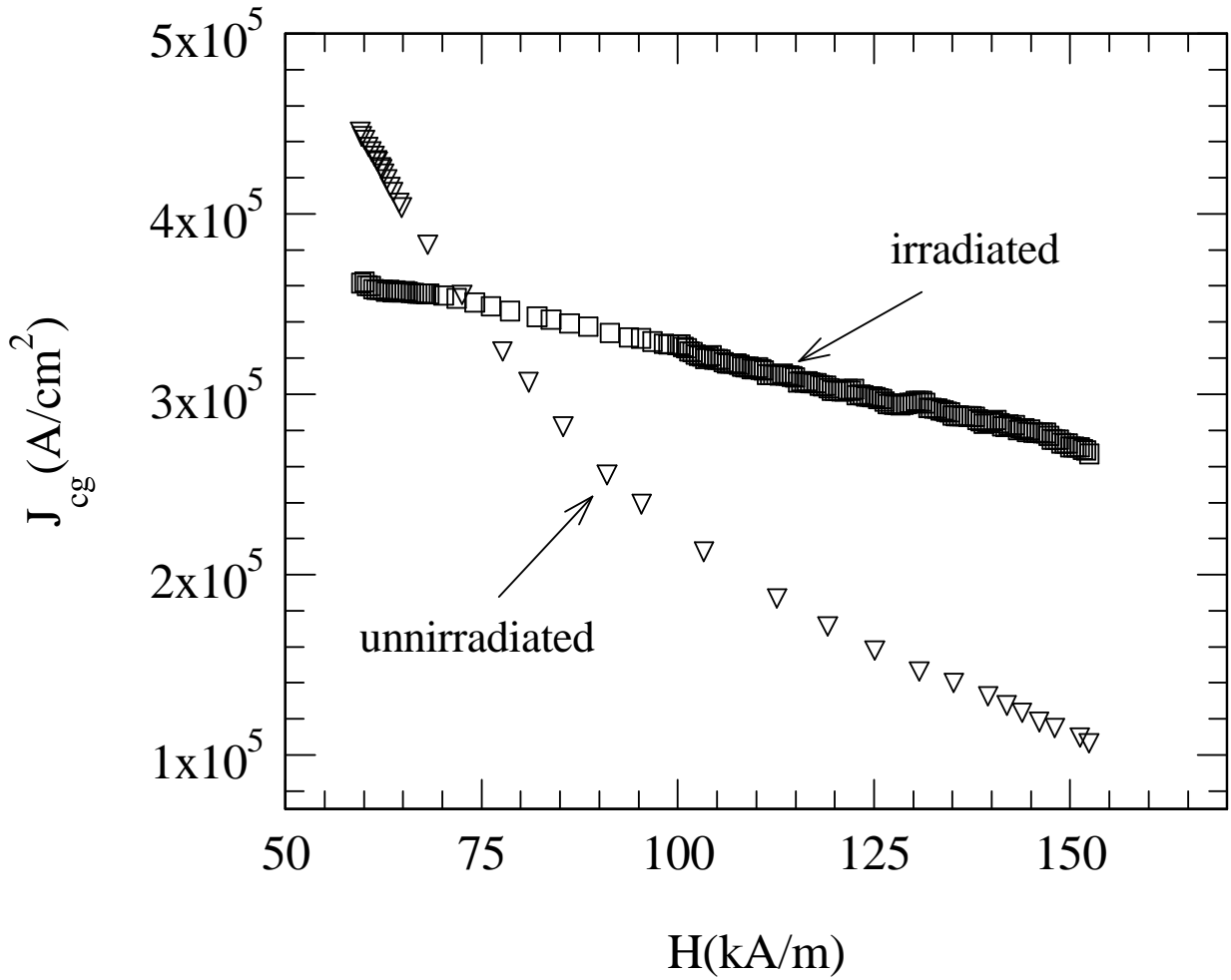

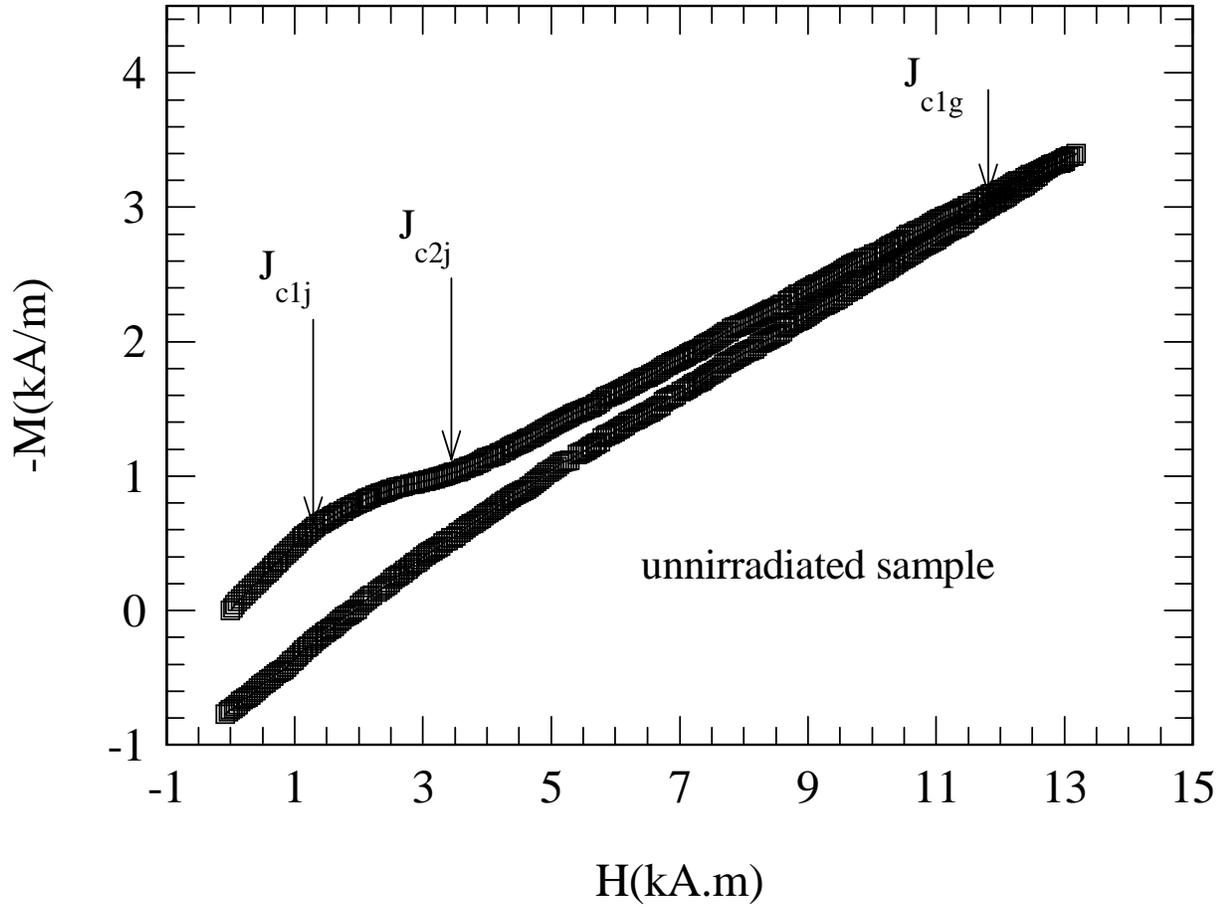

Fig. 5a

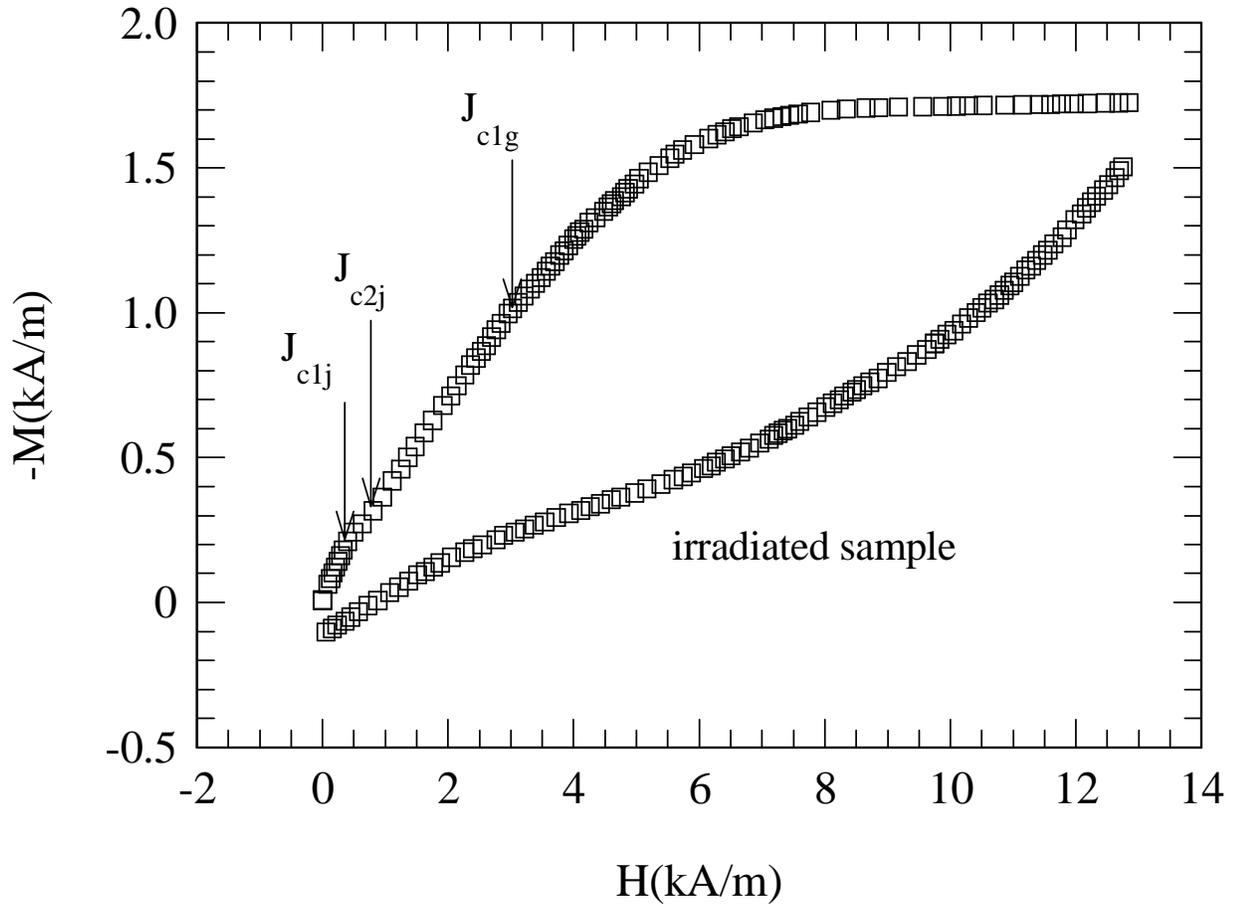

Fig. 5b